\documentclass{egpubl}
\usepackage{booktabs} % For formal tables
\usepackage{textcomp}
\usepackage{amsmath}
\usepackage{amsfonts}
\usepackage{bm}
\usepackage{url}
\usepackage{multirow}
\usepackage{multicol}
\usepackage{mathrsfs}

\DeclareMathOperator*{\argmin}{arg\,min}
\newcommand{\red}[1]{\textcolor{black}{#1}}

\JournalSubmission    % uncomment for submission to Computer Graphics Forum
\usepackage[T1]{fontenc}
\usepackage{dfadobe}  

\usepackage{cite}  % comment out for biblatex with backend=biber
\BibtexOrBiblatex
\electronicVersion
\PrintedOrElectronic
\ifpdf \usepackage[pdftex]{graphicx} \pdfcompresslevel=9
\else \usepackage[dvips]{graphicx} \fi

\usepackage{egweblnk} 
\title[Locality-Preserving Free-Form Deformation]%
      {Locality-Preserving Free-Form Deformation}

\author[T. Fukusato \& A. Maejima]
{
    \parbox{\textwidth}{\centering Tsukasa Fukusato$^{1}$\orcid{0000-0002-5090-1443}
    \hspace{2mm} Akinobu Maejima$^{2, 3}$\orcid{0000-0002-8005-9218
} \hspace{2mm} Takeo Igarashi$^{4}$\orcid{0000-0002-5495-6441}} \\
    {\parbox{\textwidth}{\centering $^1$~Waseda University, the School of Fundamental Science and Engineering, Japan
    \\$^2$~OLM Digital Inc, Japan
    \\$^3$~IMAGICA GROUP Inc, the Advanced Research Group, Japan
    \\$^4$~The University of Tokyo, the Department of Creative Informatics, Japan}
    }
}

\begin{document}

\maketitle
%-------------------------------------------------------------------------
\begin{abstract}
This paper proposes a method to estimate the locations of grid handles in free-form deformation (FFD) while preserving the local shape characteristics of the 2D/3D input model embedded into the grid, named locality-preserving FFD (lp-FFD). Users first specify some vertex locations in the input model and grid handle locations. 
The system then optimizes all locations of grid handles by minimizing the distortion of the input model's mesh elements.
The proposed method is fast and stable, allowing the user to directly and indirectly make the deformed shape of mesh model and grid. 
This paper shows some examples of deformation results to demonstrate the robustness of our lp-FFD.
\red{In addition, we conducted a user study and confirm our lp-FFD's efficiency and effectiveness in shape deformation is higher than those of existing methods used in commercial software.}

%-------------------------------------------------------------------------
%  ACM CCS 1998
%  (see https://www.acm.org/publications/computing-classification-system/1998)
% \begin{classification} % according to https://www.acm.org/publications/computing-classification-system/1998
% \CCScat{Computer Graphics}{I.3.3}{Picture/Image Generation}{Line and curve generation}
% \end{classification}
%-------------------------------------------------------------------------
%  ACM CCS 2012
%   (see https://www.acm.org/publications/class-2012)
%The tool at \url{http://dl.acm.org/ccs.cfm} can be used to generate
% CCS codes.
%Example:
\begin{CCSXML}
<ccs2012>
<concept>
<concept_id>10010147.10010371.10010387</concept_id>
<concept_desc>Computing methodologies~Graphics systems and interfaces</concept_desc>
<concept_significance>500</concept_significance>
</concept>
</ccs2012>
\end{CCSXML}

\ccsdesc[500]{Computing methodologies~Graphics systems and interfaces}
\printccsdesc   

\end{abstract}  

%-------------------------------------------------------------------------
\section{Introduction}
\label{sec:introduction}
\begin{figure}[t]%b
\centering
\includegraphics[width=1.0\linewidth]{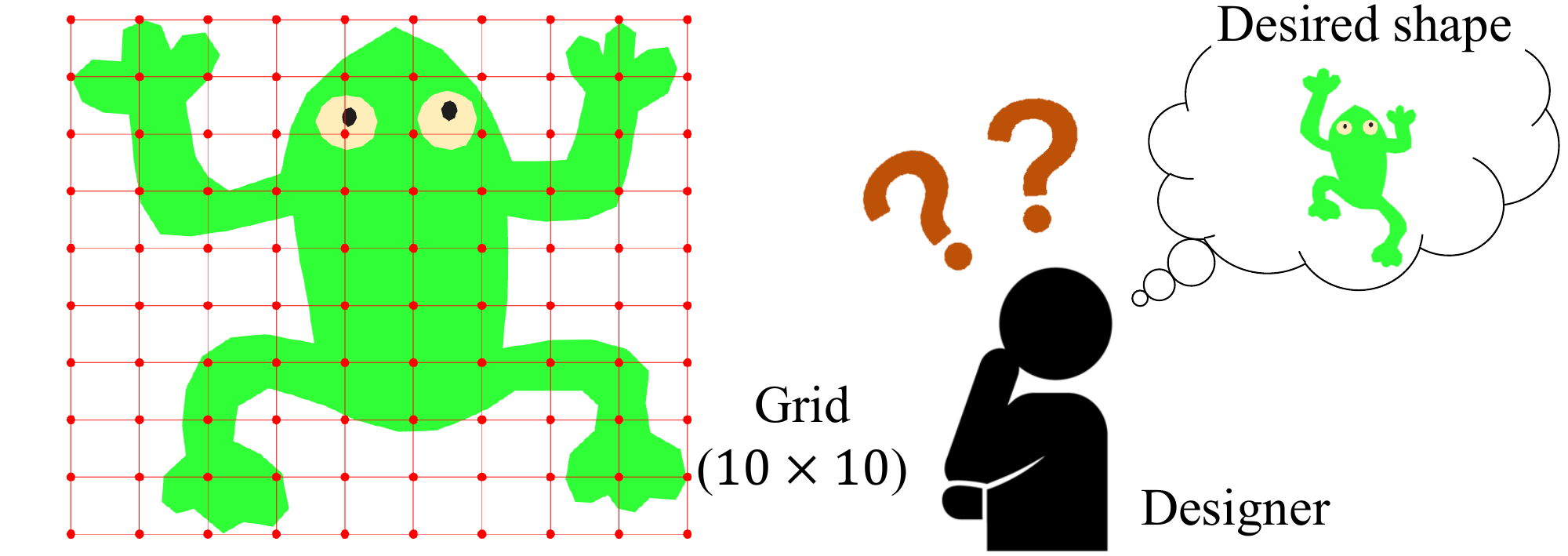}
\caption{The problem with FFD. The user must indirectly deforms vertices on the input model by manipulating grid handles (red points). The model is designed with reference to Frog from \cite{igarashi2005rigid}.}
\label{fig:limitation}
\end{figure}

Free-form deformation (FFD)~\cite{sederberg1986ffd, maccracken1996free} is a popular way to smoothly deform 2D/3D models, and is widely applied in well-known commercial software, for example, Adobe AfterEffect Mesh Warp\footnote{\url{https://www.adobe.com/products/aftereffects.html}}, Adobe Photoshop Warp Grid\footnote{\url{https://www.adobe.com/products/photoshop.html}}, Blender Lattice Modifier\footnote{\url{https://www.blender.org/}}, and Clip Studio Paint Mesh Transformation\footnote{\url{https://www.clipstudio.net/en/}}. 
In 2D cases, the user first places an $M \times N$ regular grid (e.g., an axis-aligned bounding box) on an input model. The system then parameterizes each vertex of the input model into the grid and computes each location $\vec{v} \in \mathbb{R}^{2}$ while moving the grid handles $\vec{P}_{mn} \in \mathbb{R}^{2}$ as follows:
% (see \autoref{fig:ffd})
%
\begin{equation}
\vec{v}(u,v) = \sum_{m}^{M}\sum_{n}^{N} B_{m}^{M}(u) B_{n}^{N}(v) \vec{P}_{mn}
\label{eq:ffd}
\end{equation}
\noindent
where $M$ and $N$ are the number of grid handles and the Bernstein polynomial $B_{a}^{b}(x)$ is given by 
\begin{equation}
B_{a}^{b}(x) = 
\begin{pmatrix}
b\\
a
\end{pmatrix} 
x^{a} (1.0-x)^{b-a}
\end{equation}
Since FFD has a fast computation speed and geometric continuity (e.g., $C^1$ and $C^0$ features), FFD can be used in various tasks such as texturing characters' garments in a photograph and transferring a deformation for one model to other objects. %is not model-specific 
However, standard FFDs are unsuitable for specifying certain vertex locations in the input model since users can only manipulate the grid handles (i.e., indirect manipulation), as shown in \autoref{fig:limitation}. 
Then, Hsu et al.~\cite{hsu1992ffd} have developed a system which allows users to directly manipulate the vertices on the model based on the inverse of \autoref{eq:ffd},  but they ignore characteristics of the input model. 
To make good deformations, users must add many constraints for vertex locations or grid handles (details are given in \autoref{sec:result}). 

%----------------------------
%----------------------------

\begin{figure*}[t]
\centering
\includegraphics[width=\linewidth]{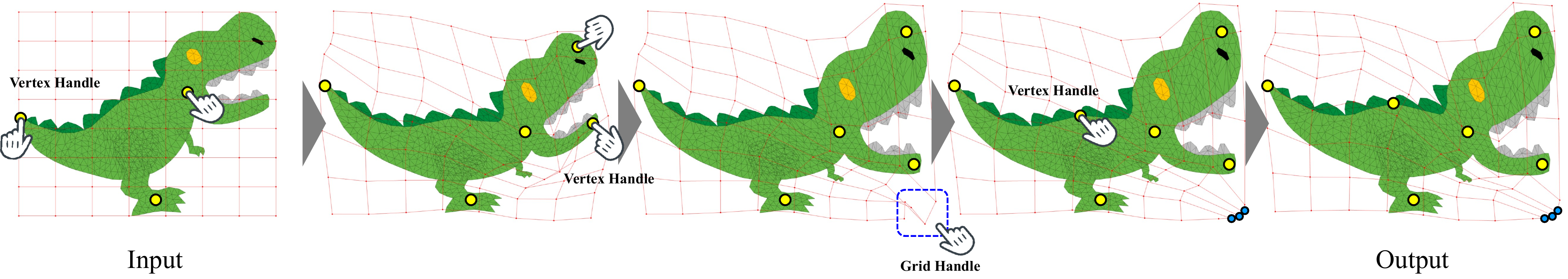}
\caption{Workflow of our proposed method. The user first places a coarse regular grid (red) on the input mesh model. The user can iteratively manipulate the vertex handles (yellow points) and the FFD grid handles (blue points) while preserving the local shape characteristics of the input model until the grid and model are deemed complete ($\#P = 8\!\times\!8$). The dinosaur model is from \cite{fukusato2023slider}.}
\label{fig:lpFFD}
\end{figure*} 
%----------------------------

Against this background, several approaches that deform an input model based on some vertex specifications without a grid, called Differential Geometry Processing (DGP)~\cite{igarashi2005rigid, sorkine2004laplacian, sorkine2007asap}, have been proposed. 
%Against this background, several approaches that deform an input model by directly specifying certain the vertex locations on the model without a grid, called Differential Geometry Processing (DGP)~\cite{igarashi2005rigid, sorkine2004laplacian, sorkine2007asap}, have been proposed. 
Their advantage is to make deformed results that preserve local details of the input model, such as Laplacian vectors and triangle shapes, and DGPs have been available on commercial software (e.g., After Effects Puppet Tool). However, their calculation depends on the number of the input model's vertices, so we need to simplify the input model~\cite{Lie2023surface} for speeding up its calculation. 
Also, the deformed results by DGP are not even $C^{1}$ continuous unlike FFD, and contain noticeable discontinuities on mesh edges when texture-mapping. 
%Also, deformation effects by DGP is generally model-specific. 
%That is, it remains difficult to reuse the deformation for one model across different objects (including images). 
In summary, DGP does not fundamentally solve the FFD problems described above. 

One solution for FFD is to estimate the locations of grid handles based on the deformed models with DGP, as an ``inverse'' problem~\cite{noh2021inverse}.
However, this approach often produces distorted mesh shapes, where the estimated grid is not suitable for user editing. The reasons are that DGP is difficult to apply to mesh models consisting of multiple isolated meshes, and FFD does not consider the shape characteristics of the input model embedded in the regular grid. 
In other words, the deformed results gained by using DGP are very different from those that can be generated using FFD, so DGP and FFD cannot be switched freely (details are given in \autoref{sec:result}). 
Furthermore, this approach requires two-step process with (i)~the DGP optimizer and (ii)~the inverse FFD optimizer. Even if the number of location constraints on the vertices of the input models is small, the optimizers are dependent on the resolution of the input model; hence, this calculation is not suitable for interactive use.

In this paper, we propose a linear optimizer that incorporates the FFD mechanism into a DGP framework for estimating locations of grid handles from a small number of vertex manipulations while preserving the local details of the input model, named locality-preserving free-form deformation (lp-FFD), as shown in \autoref{fig:lpFFD}.  
\red{We evaluated the proposed method based on visual comparison and existing distortion measures~\cite{wang2016arap++}, and found that the proposed method can better preserve localities. 
We also confirmed the intuitiveness of the proposed method through a user study.}
The proposed method can pre-compute matrices representing the local details and the FFD parameterization, making it suitable for interactive use. 
Our system allows users to export the deformed grids, making it possible to reuse them for various tasks, such as image manipulation and transferring the deformation effects. Our core idea can also be applied to other subspace deformers in 2D and 3D.
% such as cage-based and skeletal-based

\section{Related Work}
\label{sec:related}
This section reviews prior work on frameworks for (1)~FFD and (2)~DGP and describes what makes our work different. 

\subsection{Free-form Deformation}
FFDs~\cite{sederberg1986ffd, maccracken1996free} have been thoroughly investigated, and many improved versions have been developed, such as triangle sets-based~\cite{kobayashi2003tffd} and point-based types~\cite{mcdonnell2007pb}. The common feature of these methods is that the input models are parameterized in advance, so the deformation qualities are dependent on the parameterization and the amount of handles $P$ must be increased to perform detailed deformations, which is labor-intensive (e.g., 100~$\times$~100 grids). 

Some methods have been proposed to estimate the positions of handles based on sparse user-specified constraints~\cite{hsu1992ffd, liu2004near, ma2014foldover}. 
In the 2D context, L{\'e}vy~\cite{levy2001constrained} proposed a method to interactively build correspondence points between input drawings (or photos) and 2D texture patterns, and generate textured results. This method employs a conjugate gradient method, which is fast enough for interactive operations. In addition, Kraevoy et al.~\cite{kraevoy2003matchmaker} and Seo et al.~\cite{seo2010constrained} extended L{\'e}vy's method to improve the quality of texture mapping by setting additional constraints, such as gradient properties. While these methods enable users to produce smoother results, they are unsuitable for including the shape characteristics of input models. In the 3D context, Hirota et al.~\cite{hirota1999fast} incorporated a physics-based idea (a mass spring network) into FFD to preserve one shape characteristic, the volume of the input model, during grid handle operation. 
However, even though the original FFD is linear, the volume constraint is nonlinear, forcing the use of a nonlinear optimizer and making the calculation unstable.

We therefore propose a linear optimizer to stabilize the computation of grid handle estimations.

\subsection{Differential Geometry Processing}
DGP is a method that focuses on the connection of the input models' vertices $\vec{v}_i$ and estimates the position coordinates of each vertex $\vec{v}'_i$ directly by minimizing the distortion of one-ring neighborhoods (or triangle shapes) before and after the deformation. For example, Sorkine et al.~\cite{sorkine2007asap} defined the following energy function $E_{dgp}$:
\begin{eqnarray}
E_{dgp} = \sum_{i = 1}^{M}\sum_{ j \in \mathscr{L}(i)} \|(\vec{v}'_{i} - \vec{v}'_{j}) - R_{i} (\vec{v}_{i} - \vec{v}_{j}) \|^{2}
\label{eq:dgp}
\end{eqnarray}
\noindent
where $M$ is the number of vertexes, $\mathscr{L}(i)$ is the set of all vertices connected with the $i$-th vertex, and $R_{i}$ is the approximate rigid transformation that fits the edges. Please see the detail of existing method. 

Since DGP can be applied to other shape characteristics (e.g., triangle shapes~\cite{igarashi2005rigid, alexa2000rigid, baxter2008rigid, baxter2009compatible, baxter2009n} and strokes~\cite{whited2010betweenit, fukusato2016active}), various linear/nonlinear constraints (e.g., length/area/volume~\cite{jin2014deformation} and injective mapping~\cite{schuller2013locally}), and some applications (e.g., blending weight computation~\cite{jacobson2011bounded} and collision handling~\cite{muller2015air}) have been proposed. 
However, since these methods have been studied separately from the FFD approaches, their results (which have been studied in the past) have not been applied in the field of FFD. 
%As a result, artists who use FFD tools to texture cartoon characters and design CAD models are still unable to benefit from them and must still resort to using basic FFD tools.
In addition, these methods may become slow since the number of variables tends to be larger than that of the grid handles. 
And, unlike FFD, the deformed results by DGPs are not $C^1$ continuous, so the lack of smoothness is clearly visible when applying a texture mapping algorithm to them. 
%And, unlike FFD, DGPs are model-specific, so the artist's effort put into obtaining a desired deformation for one model cannot be reused across different models.

Therefore, we bridge DGP and FFD techniques by estimating locations of FFD handles while minimizing the distortion of the input model's shape. Taking a somewhat similar approach, Huang et al.~\cite{huang2006subspace} optimized cage shapes (indirect deformer) based on characteristics of the input model. However, since their goal was to speed up the DGP process under nonlinear constraints, they did not discuss the qualities of the estimated cage shape and it remains difficult for users to manipulate the cage. In addition, since their algorithm is nonlinear, the solution depends on the number of iterations and initial conditions. 
It remains difficult for users to iteratively edit the locations of grid handles, as in texture mapping process~\cite{noh2021inverse, fukusato2022interactive}. 

Based on their work, we attempt to estimate a natural-looking grid shape in FFD methods by using linear approximation to make it easier for users (including amateur artists and novices) to iteratively manipulate the grid and mesh models. %results.
\section{Method}
\label{sec:method}
Since an FFD (see \autoref{eq:ffd}) is linear, a weight matrix $W \in \mathbb{R}^{M \times N}$ can be constructed to compute each vertex location in the input model $V = (\vec{v}_{1}, \vec{v}_{2}, \cdots, \vec{v}_{M})^{T}$ based on the locations of the grid handles $P = (\vec{P}_{1}, \vec{P}_{2}, \cdots, \vec{P}_{N})^{T}$ by assigning coefficient values of the Bernstein polynomials. Note that to simplify the FFD equation, the index of each grid handle in FFD ($= width \times height$) is represented as a single variable.
% \times depth
%
\begin{equation}
\begin{pmatrix}
\vec{v}_{1}\\
\vdots\\
\vec{v}_{M}
\end{pmatrix}
= W
\begin{pmatrix}
\vec{P}_{1}\\
\vdots\\
\vec{P}_{N}
\end{pmatrix}
\label{eq:weight}
\end{equation}
\noindent
Note that the matrix~$W$ can be replaced with other linear blending, such as a barycentric energy that reflects the relative position of a vertex in a specific cell.

We then re-visit the DGP formulas $E_{dgp}$ (see \autoref{eq:dgp}). These methods are based on linear optimization to find the mesh vertex locations $V' = (\vec{v}'_{1}, \vec{v}'_{2}, \cdots, \vec{v}'_{M})^{T}$ that locally minimize distortions under given handle constraints. Therefore, by substituting \autoref{eq:weight} into $V'$ in \autoref{eq:dgp}, we define a new problem that solves the locations of the grid handles~$P'$ while integrating shape characteristics of the input model, as follows:
\begin{eqnarray}
 E_{ml} 
 &=& \|L V' - T L V \|^{2}\nonumber \\
 &=& \|(L W) P' - T L V\|^{2}
\end{eqnarray}
\noindent
where $L \in \mathbb{R}^{M \times M}$ is the Laplacian matrix computed from the input model and $T$ is a matrix that stores a set of approximated rigid transformation matrices of one-ring neighbors~\cite{sorkine2007asap}. Please refer to previous studies for more details.
Note that it is possible to apply this idea to other DGP contexts, such as triangle-based~\cite{igarashi2005rigid, baxter2009compatible} and stroke-based schemes~\cite{fukusato2016active}.

In addition, as with Hsu et al.~\cite{hsu1992ffd}, when specifying the locations of a few mesh vertices directly, this vertex setting can be considered as location constraints, named vertex handles, by using the variables of grid handles~$P'$. Let $C = \{\vec{c}_{i}\}$ be the set of user-specified vertex handles (= target locations); the vertex handle constraint term is represented as follows:
%moving the mesh vertices 
%
\begin{eqnarray}
 E_{mp} 
 &=& \sum_{i \in C}\|\vec{v}'_{i} - \vec{c}_{i}\|^{2} \nonumber \\
 &=& \sum_{i \in C}\|\vec{W}_{i} P' - \vec{c}_{i}\|^{2}
\end{eqnarray}
where $\vec{W}_{i} \in \mathbb{R}^{N}$ is the $i$-th weight row vector of $W$ for computing the location of the $i$-th mesh vertex. Note that since our system is based on an FFD scheme, vertex constraints can also be added to arbitrary locations in the regular grid.

\begin{figure}[t]%b
\centering
\includegraphics[width=\linewidth]{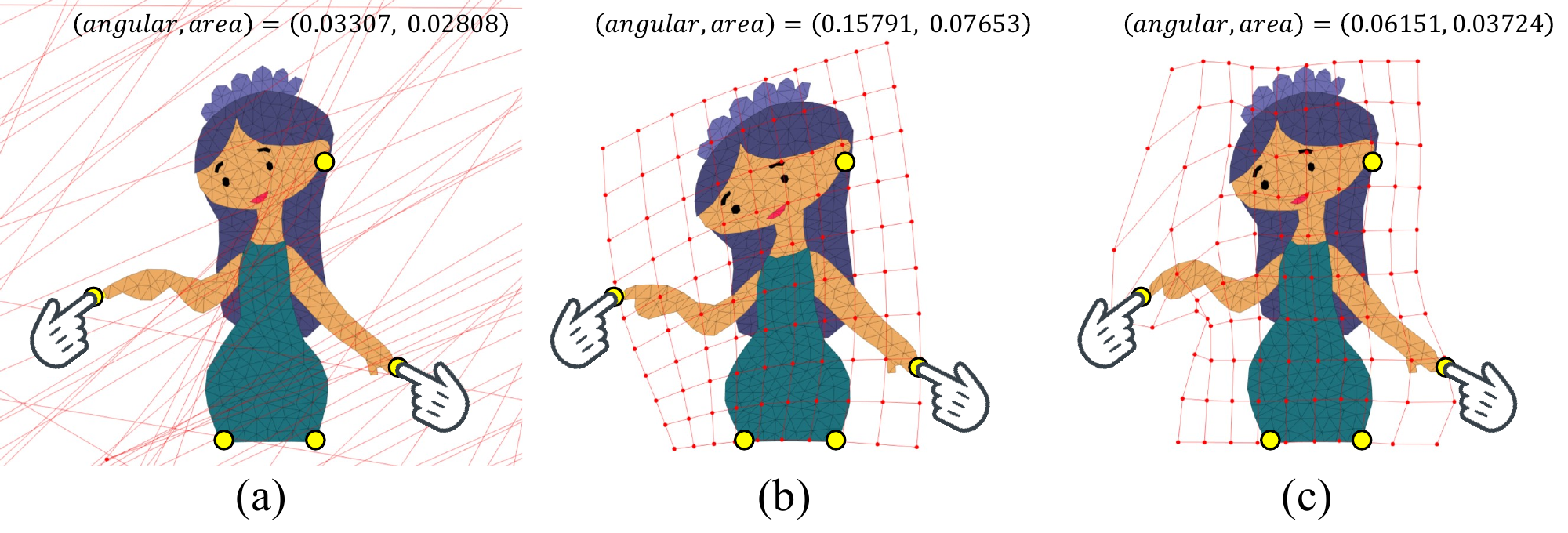}
\caption{The comparison of regularization methods. The deformed results were obtained (a)~without any regularization, (b)~with Noh et al.~\cite{noh2021inverse}, and (c)~with $E_{gr}$ ($\#P = 10\!\times\!10$). The angular and area distortions are computed by \cite{wang2016arap++}. The model is designed with reference to Girl from \cite{sykora2009rigid}.}
%horizontal/vertical edges-based
%\caption{The impact of the regularization term. The deformed result was obtained (a)~without the regularization, (b)~horizontal/vertical edges-based~\cite{noh2021inverse}, and (c)~with $E_{gr}$ ($\#P = 10\!\times\!10$). The model is designed with reference to Girl from \cite{sykora2009rigid}.}
\label{fig:damping}
\end{figure} 
%example-image-duck

Similarly, for manipulating some grid handles (i.e., indirect manipulation), we add a constraint term of user-specified grid handle locations $D = \{\vec{d}_{i}\}$ as follows:
\begin{eqnarray}
 E_{gp} 
 &=& \sum_{i \in D}\|\vec{P}'_{i} - \vec{d}_{i}\|^{2}
\end{eqnarray}
However, the above terms are not sufficient for user editing since the estimated locations of grid handles may significantly stick out of the screen even if natural-looking mesh shapes can be obtained (\autoref{fig:damping}(a)). 
%Note that this problem (unreasonable location) also occurred in existing subspace deformers~\cite{huang2006subspace}, but it was not well-discussed in their paper. 
%%in Huang et al.~\cite{huang2006subspace},
Noh et al.~\cite{noh2021inverse} have examined a smoothing term based on the relative location relationship between grid handles, but often make stretched (or shrinked) shapes (\autoref{fig:damping}(b)). In this paper, we add a simple regularization term $E_{gr}$ as constraints on grid handles not edited by users:
%, inspired by \cite{lewis2010direct}:
%To address this issue,
%
\begin{eqnarray}
 E_{gr} &=& \sum_{i \notin D}\|\vec{P}'_{i} - \vec{P}_{i}^{0}\|^{2}
 \label{eq:regularization}
\end{eqnarray}
where $\vec{P}_{i}^{0}$ denotes the initial position of the vertices of the regular grid. \autoref{fig:damping}(c) shows an example of the effect of this constraint. \red{From the distortion measures~\cite{wang2016arap++}, we confirm $E_{gr}$ is still simple but can balance grid handles and local detail preservation.}

To balance between preserving the input mesh’s local detail and setting vertex/grid handle locations, we define the following energy function:
\begin{eqnarray}
 \argmin_{P'}\{\lambda_{ml} E_{ml}  + \lambda_{mp}E_{mp} + \lambda_{gp}E_{gp} + \lambda_{gr}E_{gr}\}
 \label{eq:opti}
\end{eqnarray}
\noindent
where $\lambda_{ml}$, $\lambda_{mp}$, $\lambda_{gp}$, and $\lambda_{gr}$ are a weight values (we empirically set them as $\lambda_{ml} = 1.0$, $\lambda_{mp} = 1.0e+02$, $\lambda_{gp} = 1.0e+02$, and $\lambda_{gr} = 1.0e-02$, respectively).

As in the existing DGP techniques~\cite{sorkine2007asap, levi2014smooth}, we iteratively optimize \autoref{eq:opti}. At each iteration, we compute the matrix $T$ based on the intermediate shape, solve the locations of the grid handles $P'$ with LU decomposition, and employ a forward FFD~(see \autoref{eq:weight}) to update the model shape based on the estimated grid handles.
%(i.e., the intermediate result).

The advantage of our lp-FFD is that the local shape characteristics of the input mesh model embedded into the 2D/3D regular grid can be accounted for, but the locations of the grid handles $P'$ can be optimized. That is, lp-FFD makes it easy to interactively repeat the direct manipulation of vertex handles and the indirect manipulation of grid handles (see \autoref{fig:lpFFD}). 
%The advantage is that the proposed method considers shape characteristics but only optimizes the locations of the grid point $P'$, so lp-FFD is easy to interactively repeat manipulating grid points and vertex handles (see \autoref{fig:lpFFD}). 
In addition, the number of variables in the formula can be the number of grid handles in FFD only and $W$ can be pre-computed, like the Laplacian matrix~$L$, making it suitable for interactive uses.
%can be reduced to the number of grid handles

\section{Results}  
\label{sec:result}
\begin{figure}[t]
\centering
\includegraphics[width=1.0\linewidth]{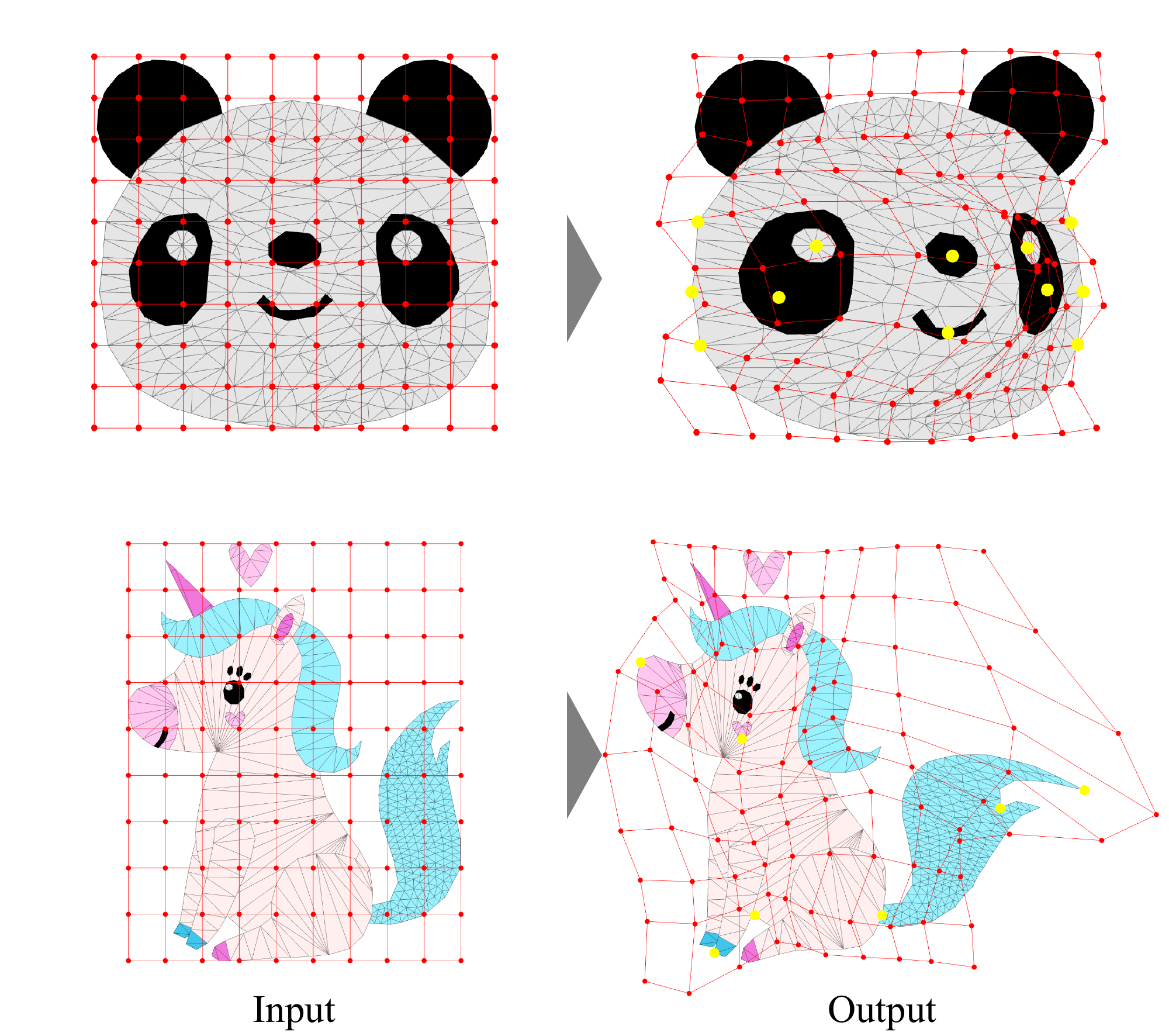}
\caption{Examples of deforming 2D layered models with lp-FFD ($\#P = 10\!\times\!10$). The models are from \cite{fukusato2022vdf} and \cite{fukusato2023slider} respectively.}
\label{fig:2D}
\end{figure}

\begin{figure}[t]
\centering
\includegraphics[width=1.0\linewidth]{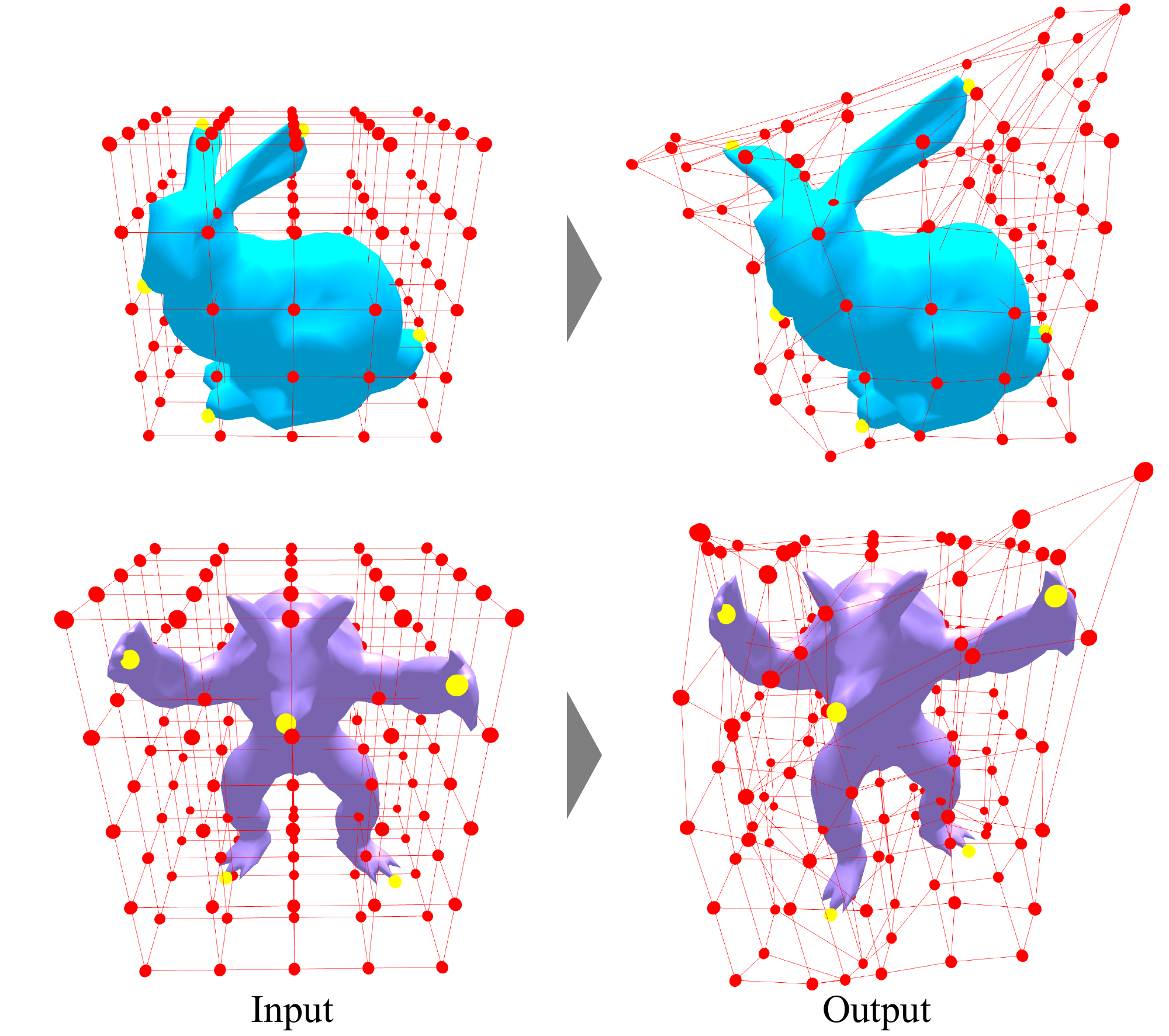}
\caption{Examples of deforming the Stanford Bunny ($\#V = 502$) and the Stanford Armadillo ($\#V = 1502$) with lp-FFD ($\#P = 5\!\times\!5\!\times\!5$).}
\label{fig:3D}
\end{figure}

We build on animation authoring tools~\cite{schuller2013locally, fukusato2022vdf} and add several functions to deform the input model. Our prototype system was implemented on a $64$-bit Windows $11$ laptop (Intel\textcircled{\scriptsize R}Core$^{TM}$ i$7$-$8500$U CPU@$1.50$GHz and $16.00$GB RAM) using standard OpenGL and GLSL. \autoref{fig:2D} and \autoref{fig:3D} show some examples of deforming 2D/3D models (with estimated grid handles) using our lp-FFD. 

%----------------------

%----------------------

%---------------------------
\begin{figure}[t]
\centering
\includegraphics[width=1.0\linewidth]{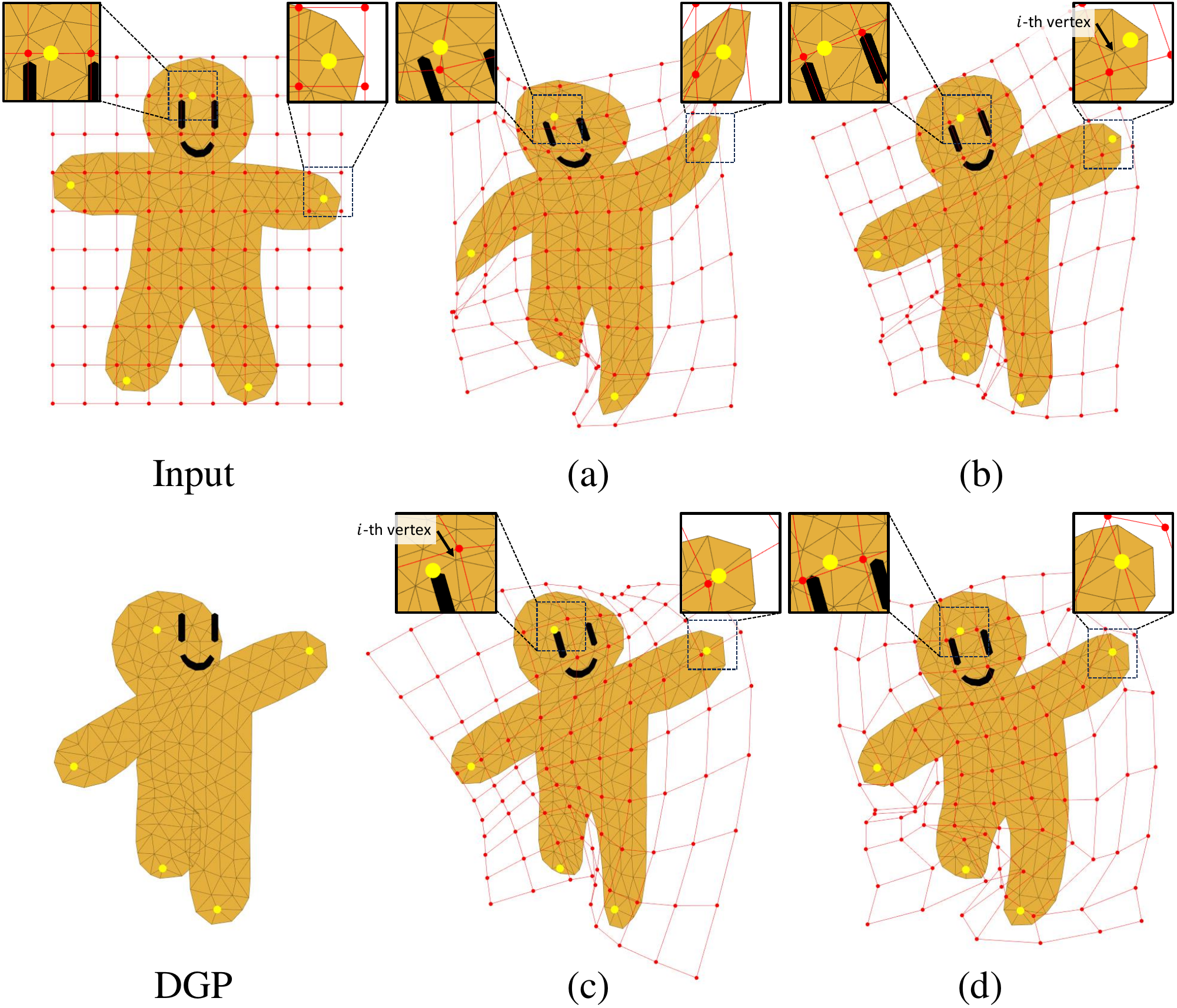}
\caption{Comparing a deformed model and grid ($\#P = 10\!\times\!10$) using (a)~Hsu et al.~\cite{hsu1992ffd}, (b)~Schaefer et al.~\cite{schaefer2006image}, (c) DGP~\cite{sorkine2007asap} + inverse FFD~\cite{noh2021inverse}, and (d)~lp-FFD. The model was designed with reference to Ginger Man from \cite{igarashi2005rigid}.}%+ the regularization term
\label{fig:comparison}
\end{figure}
%---------------------------

\subsection{Comparison of Shape Deformation}  
\label{sec:mesh_comparison}
\red{Various shape deformation methods (e.g., point-based and skeletal-based) have been proposed, but for comparison, it is necessary to select appropriate existing methods. In this paper, we selected the following three deformation methods that can be used to estimate the locations of grid handles in FFD and} applied both the proposed method (lp-FFD) and some existing methods to a 2D layered model, ``Ginger Man'' consisting of right and left eyes, a mouth, and body parts (i.e., triangulated meshes). 
To quantify \red{how well these methods can preserve local details after deforming mesh models~$V'$}, we compute the angular and area distortion measures, as defined in \cite{wang2016arap++}.
%To quantify the distortion measures of deformed mesh models~$V'$, we use the angular and area distortions, as defined in \cite{wang2016arap++}.

The first method is a direct manipulation technique proposed by Hsu et al.~\cite{hsu1992ffd}, plus the regularization term (see \autoref{eq:regularization})  for stable computation. 
We denote this method's idea is similar to Lewis et al.~\cite{lewis2010direct}, which estimates the parameters in blendshape deformers under vertex constraints. As shown in \autoref{fig:comparison}(a) and \autoref{tab:distortion}, the deformed result was unnatural but the area distortion score became low. A possible reason is that the areas affected by vertex handle manipulations are small and the undeformed areas (e.g., the Ginger Man's body) reduced the score. Also, this method cannot preserve local details of the input model embedded into the grid, so the angular distortion score was high. Therefore, it is thought that many additional constraints are needed to make good deformation effects.

The second method is a 2D grid deformer based on Moving Least Squares~\cite{fukusato2022interactive, schaefer2006image}. It can deform a 2D grid while accounting for the rigidity of the regular grid, but it does not take into account any characteristics of the input model embedded into the grid. As a result, applying the estimated locations of grid handles to the input model may produce distorted deformations, such as mesh vertices being far from user-specified locations (yellow) and a thinning of the Ginger Man's leg (see \autoref{fig:comparison}(b)), and both angular and area distortions tend to be high, as shown in \autoref{tab:distortion}.

The third method is to (i)~deform input models using DGP and then (ii)~estimate the locations of grid handles~$P'$ using inverse FFD~\cite{noh2021inverse}. In the original inverse FFD paper, input models were generated using Hashimoto's method~\cite{hashimoto2020neurally} (which is based on a DGP technique), and a baricentric energy that reflects the relative vertex position in a particular cell and its four control points were used for FFD. 
In this comparison, to match the conditions with ours, we used Sorkine's DGP (see \autoref{eq:dgp}) in the first step and FFD (see \autoref{eq:ffd}) in the second step. %\autoref{fig:comparison}(c) shows an example of estimating the grid handles. 
From \autoref{fig:comparison}(c) and \autoref{tab:distortion}, we can see that this approach outputs distorted mesh shapes and both distortion measures are high. We denote that the user-specified locations of grid handles are far from the corresponding mesh vertices in the deformation results.
%From this result, we can see that this approach outputs a distorted mesh shape and a grid shape unsuitable for editing. In addition, both distortion measures are high (see \autoref{tab:distortion}). 
The main reason is that DGP is not suitable for deforming a mesh model consisting of multiple isolated meshes, so the deformation results with DGP are not suitable as input data for solving the inverse problem. 
In addition, the inverse FFD forcibly estimates the grid handles $P'$ from the DGP results and outputs final deformations through FFD. However, FFD cannot take into account any shape characteristics of the input model embedded in a regular grid. 
In summary, the shapes that can be generated in the first stage (DGP) and in the second stage (FFD) are different, making it unsuitable for both grid and vertex handle manipulations.
Moreover, this approach requires the solution of two optimizations (i.e., solving mesh vertex $V'$ and grid handles $P'$), and the total computation speed tends to increase according to the number of vertices on the input mesh model even if the grid size is small (see \autoref{tab:ctime}).

 %------------------------------
\begin{table}[t]
\centering
 \caption{Comparison of angular distortion and area distortion measures on the Ginger Man model ($\#P = 10\!\times\!10$).}
 \begin{tabular}{l|c|c}
     \hline
     Method & Angular & Area \\
     \hline \hline
     Hsu et al.~\cite{hsu1992ffd} & 0.21808 & 0.14608 \\ %The simplified inverse FFD
     Schaefer et al.~\cite{schaefer2006image} & 0.21793 & 0.15529 \\
     DGP~\cite{sorkine2007asap} $+$ inverse FFD~\cite{noh2021inverse}  & 0.20432 & 0.24516 \\
     lp-FFD (Ours)  & \textbf{0.14694} & \textbf{0.11644} \\
    \hline
 \end{tabular}
\label{tab:distortion}
\end{table}
 %------------------------------

 %------------------------------
\begin{table}[t]
\centering
 \caption{The computation time using single-threaded computing [ms]. Note that matrices $L$ and $W$ were not pre-computed in this time measurement.}
 \begin{tabular}{c|c|c|cc|c}
     \hline
     \raisebox{-0.3mm}{Model} & 
     \raisebox{-0.3mm}{\#V} & %Vertex Point
     \raisebox{-0.3mm}{\#P}  & 
     \multicolumn{2}{c|}{\raisebox{-0.3mm}{\cite{sorkine2007asap} \hspace{1.7mm}$+$\hspace{1.7mm}\cite{noh2021inverse}}} & 
     \raisebox{-0.3mm}{Ours}\\
     \hline \hline
     \multirow{3}{*}{Ginger Man} &\multirow{3}{*}{225}& $5\!\times\!5$  & \multirow{3}{*}{15.6}  & 3.00 & \textbf{3.20} \\
                               &                     & $10\!\times\!10$ & & 6.80 & \textbf{6.40} \\
                               &                     & $15\!\times\!15$ & & 19.2 & \textbf{16.4} \\
     \hline
     \multirow{3}{*}{Frog}     & \multirow{3}{*}{315}& $5\!\times\!5$   & \multirow{3}{*}{23.8} & 4.00 & \textbf{4.40} \\
                               &                     & $10\!\times\!10$ & & 8.60 & \textbf{8.80} \\
                               &                     & $15\!\times\!15$ & & 26.2 & \textbf{25.8} \\
     \hline
     \multirow{3}{*}{Panda}     & \multirow{3}{*}{444}& $5\!\times\!5$   & \multirow{3}{*}{31.8} & 5.40 & \textbf{6.20} \\
                               &                     & $10\!\times\!10$ & & 15.0 & \textbf{15.0} \\
                               &                     & $15\!\times\!15$ & & 34.0 & \textbf{32.2} \\
     \hline
     \multirow{3}{*}{Unicorn}  & \multirow{3}{*}{549}& $5\!\times\!5$   & \multirow{3}{*}{34.2} & 6.60 & \textbf{6.60} \\
                               &                     & $10\!\times\!10$ & & 22.6 & \textbf{16.4} \\
                               &                     & $15\!\times\!15$ & & 42.2 & \textbf{34.0} \\
     \hline
     \multirow{3}{*}{Girl}     & \multirow{3}{*}{637}& $5\!\times\!5$   & \multirow{3}{*}{51.6} & 8.00 & \textbf{11.8} \\
                               &                     & $10\!\times\!10$ & & 31.6 & \textbf{21.8} \\
                               &                     & $15\!\times\!15$ & & 67.0 & \textbf{43.0} \\
     \hline
     \multirow{3}{*}{Dinosaur} & \multirow{3}{*}{734}& $5\!\times\!5$   & \multirow{3}{*}{69.8} & 9.40 & \textbf{15.6} \\
                               &                     & $10\!\times\!10$ & & 36.6 & \textbf{26.8} \\
                               &                     & $15\!\times\!15$ & & 81.4 & \textbf{44.2} \\
    \hline
 \end{tabular}
\label{tab:ctime}
\end{table}
 %------------------------------

In contrast, the proposed method can estimate relatively stable grid handle locations and natural-looking deformed models, as shown in \autoref{fig:comparison}(d). And, as is evident in \autoref{tab:distortion}, our method (lp-FFD) consistently gives the best balance between the angular distortion and the area distortion. In addition, lp-FFD requires only one optimization step for estimating grid handle locations, which is computationally light (see \autoref{tab:ctime}), and allows a pre-computation of each matrix. We denote when the number of grid handles is small compared to the number of mesh vertices ($M > N$), our lp-FFD has a faster computation speed compared with standard DGPs, as with Huang et al~\cite{huang2006subspace}.

%---------------------------
\begin{figure}[t]
\centering
\includegraphics[width=\linewidth]{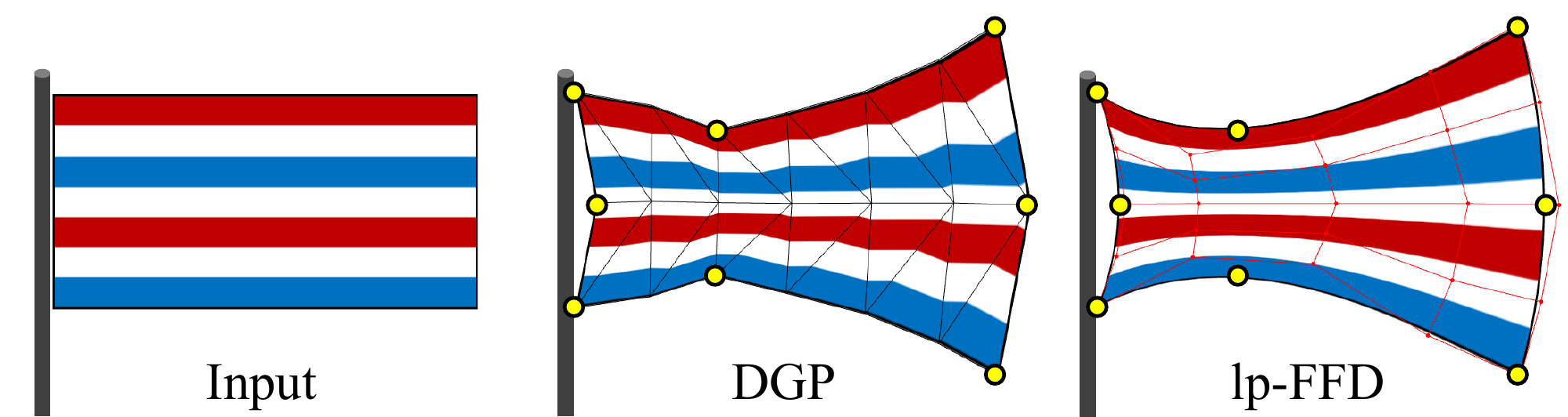}
\caption{Deforming a flag image with DGP (with linear texture mapping) and lp-FFD. The number of mesh vertices $\#V$ and grid handles $\#P$ is $21$ and $25 (= 5\!\times\!5)$ respectively, and eight vertex handles (yellow points) are attached.}
\label{fig:imageDeform}
\end{figure} 
%---------------------------

\subsection{Comparison of Image Deformation}  
\label{sec:image_comparison}
We compare image manipulations with DGP and lp-FFD. Existing approach with DGPs~\cite{igarashi2005rigid, alexa2000rigid, baxter2009compatible, baxter2009n, fukusato2022interactive, hashimoto2020neurally} first generates a triangulated mesh from an input image. Next, users deforms the mesh and maps the image to the deformed mesh. However, their results depend on triangulation and texture mapping algorithm, and often have noticeable discontinuities in the deformed image, as shown in \autoref{fig:imageDeform}(middle). 
On the other hand, the output of our lp-FFD is locations of grid handles in FFD, so users can easily obtain smoother results (i.e., $C^{1}$ continuous) based on the deformed grid even if low-resolution mesh were used (see \autoref{fig:imageDeform}(right)).

%additional point++++++++++++++++++++++++++++
%http://norawillett.com/secondary_animation/data/paper.pdf
\subsection{User Study}  
\label{sec:user_study}
To investigate the effectiveness of our method (lp-FFD), we conducted a user study where people compare the three existing methods with lp-FFD (see \autoref{sec:mesh_comparison}).

%\vspace{2mm}
\subsubsection{Procedure}
We invited 9 participants (P1, \dots, P9) aged 20s-30s (Avg $= 28.33$, SD $= 4.74$). Each participant was asked to fill out a form asking about their experience with manipulating images and 2D/3D mesh models using commercial software. 
%9
P1 (female) had extensive experience in creating 2D/3D models
with Autodesk Maya and Rhinoceros/Grasshopper ($>$ 15 years), and image manipulation tools with Adobe Photoshop ($>$ 20 years). 
%3, 5
P2 - 4 (3 male) were casual users who had at least a few year’s experience of using Blender and Adobe Photoshop. 
% ($>$ 2 years) and Adobe Photoshop ($>$ 6 years). 
P5 - 9 (3 male and 1 female) had an interest in designing images and 2D/3D mesh models but no prior experience.
%First, each participant was given a brief overview of our meshing tool by an instructor, who worked through a step-by-step tutorial (5-10 min) to familiarize them with the framework. 

We provided them with several videos recording of the authors' manipulations on the Ginger Man model (see \autoref{fig:comparison}), and asked them to visually compare the above methods. %until they were satisfied. 
For each participant, the order of the method (for example, the first is Hsu et al., the next is Schaefer et al., \dots) was randomly shuffled, and we did not tell him/her which method was lp-FFD. 
At the end of the comparison, the participants filled out a questionnaire consisting of two questions about their impressions using a seven-point Likert scale (from $1=$ ``\textit{extremely dissatisfied}'' to $7 =$ ``\textit{extremely satisfied}''). The purpose of these questions was to analyze their subjective impression. The two questions were as follows: (Q1)~``\textit{I think that manipulating pin handles was intuitive (= ease of specifying locations of mesh vertex)}'' and (Q2)~``\textit{I satisfied with the deformation results}.'' The total evaluation process took approximately 15 minutes per participant.

%\vspace{2mm}
\subsubsection{Observations and User Feedback}
\begin{figure}[t]
\centering
\includegraphics[width=\linewidth]{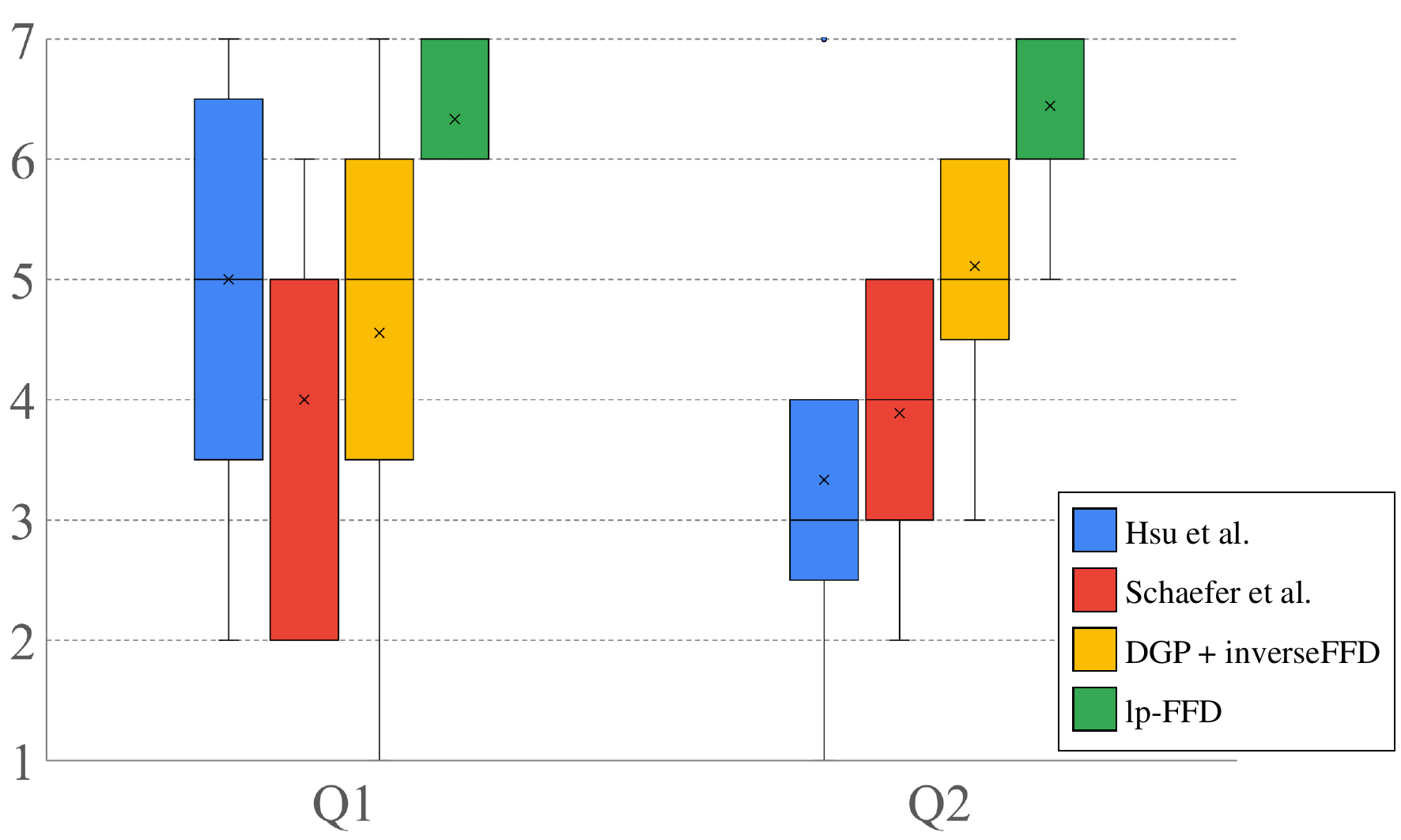}
\caption{The boxplot of participant response to the post-experiment questionnaire. The cross mark and the thick gray line are the average and the median respectively.}
\label{fig:graph}
\end{figure}

\autoref{fig:graph} shows the post-experiment questionnaire results. According to these results, the participants rated lp-FFD more positively.
%We also calculated p-value by running a Wilcoxon signed-rank test. The result was p = 2.18e-02, which is significant at p < 0.05 for a two-tailed hypothesis.

We will now discuss the comments about the reasons for this answer in more detail. When using Hsu et al.~\cite{hsu1992ffd}, some participants had some trouble finding grid handles' locations and generating the deformed model. For example, \textit{``The mesh deformation is somewhat unexpected in that the width of the arm is not maintained''} (by P1, P3, P6, P8, and P9), and \textit{``Since only the mesh near the pins (yellow) is deformed, I had to manipulate many handles to achieve the desired deformation''} (by P3). 
%When editing the input character's posture by manipulating the vertex handles, the character's arms and legs often became thinner. 
%To modify it, I had to manipulate many grid handles, so didn't think this method was convenient'' (by PX). 
From these comments, Hsu et al. enables users to easily specify locations of few mesh vertices using vertex handles ($>$ 4), but the estimated results tend to be distorted, making it difficult to use ($<$ 4). 
In the case of 2D grid deformer~\cite{schaefer2006image}, two participants answered that \textit{``I think the deformation is more global than expected, which results in unintended editing results''} (by P1, P8, and P9), \textit{``Some regions may become particularly large or small''} (by P4 and P6), and \textit{``When deforming the mesh model, it was noticeable that the deformed results do not match the location of the vertex handle''}  (by P9).
That is, it is thought that although this approach can naturally deform the 2D grid, the participants were not satisfied with the deformation results of the mesh model and preferred to control the locations of the mesh vertex using vertex handle. 
In the case of using DGP~\cite{sorkine2007asap} $+$ inverse~FFD~\cite{noh2021inverse}, four participants stated, \textit{``Compared to the method~A ($=$ Hsu et al.) and B ($=$ Schaefer et al.), the deformed shapes seem natural at a glance. But the shape of facial region was partially distorted''} (by P1, P3, and P4) and \textit{``I could not make the desired deformation since the vertex handle locations did not match the mesh vertices (i.e., point constraints)''} (by P4 and P5), and \textit{``The calculation speed is a little slow''} (by P3 and P8). 
%That is, it is thought that this approach are unsuitable for deforming the input model since its local details were lost. 
On the other hand, when using lp-FFD, several participants answered that \textit{I can directly set the vertex locations in the input mesh, and the deformed result was the most natural} (by P1, P3, P4, P8, and P9) and \textit{``The deformation results are very consistent with the original model. Compared to the method~C ($=$ DGP + inverseFFD), the process is smoother''} (by P6). In summary, it is thought that our lp-FFD not only has low distortion values and faster computation speed (see \autoref{sec:mesh_comparison}), but also allows users to design intended deformations ($>$ 6).

%-----------------------------------------------------------
\subsection{Application}  
\label{sec:application}
\begin{figure}[t]
\centering
\includegraphics[width=\linewidth]{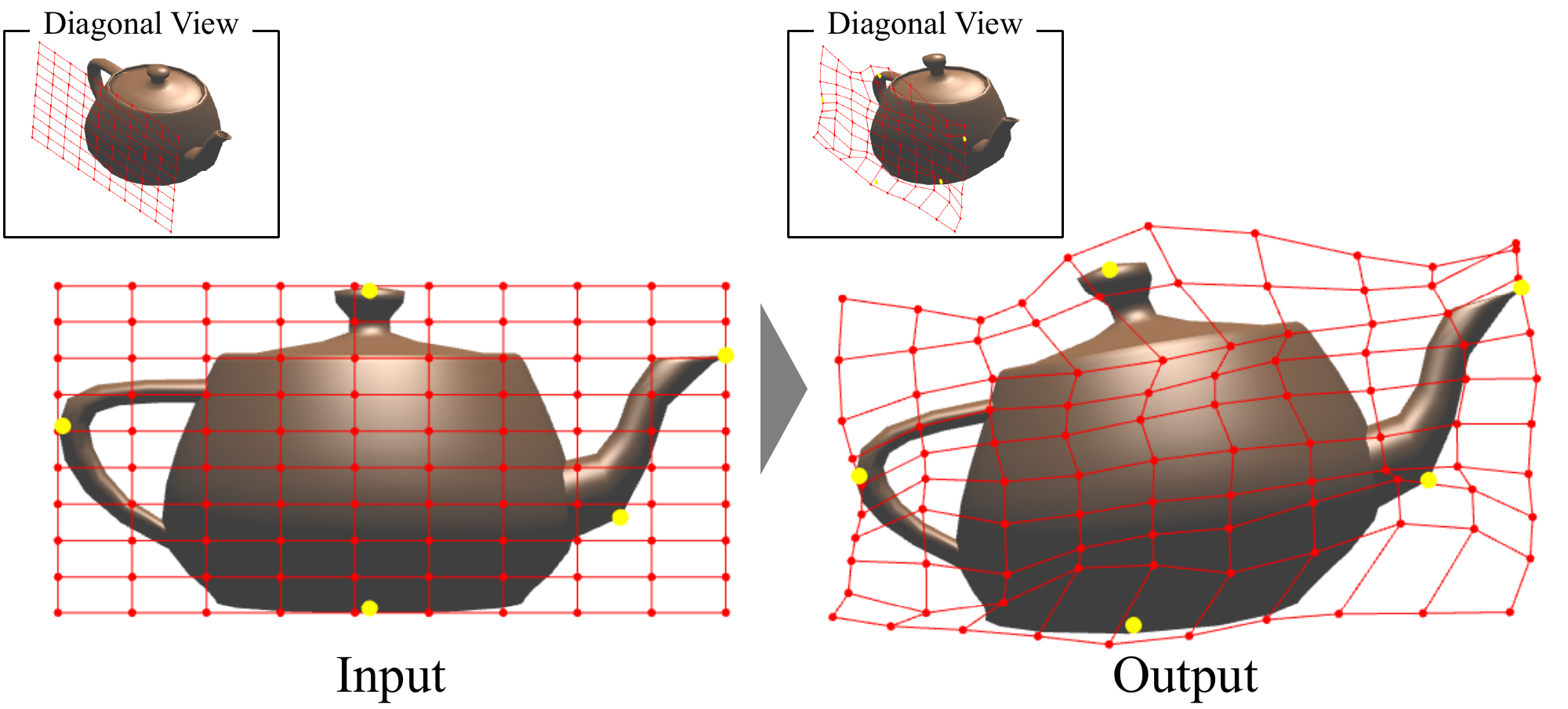}
\caption{An example of screen-space lp-FFD. By corresponding the input 3D model (the Utah Teapot: $\#V = 530$) with a 2D regular grid, the user can simply deform it through the 2D-space manipulation and optimization ($\#P = 10\!\times\!10$).}
\label{fig:application1}
\end{figure}

We investigated applications of the proposed method by users who frequently use FFD in commercial software. One group of consumate artists commented, ``\textit{In cartoon production, artists use mesh warping in Adobe AfterEffect to soften the impression of rendered 3D animations in post-processing. This idea is very simple but can create inbetweening-like effects while reducing the 3DCG-like look.}'' 
From this comment, our lp-FFD can help design such effects by corresponding 3D scenes with the 2D screen, as shown in \autoref{fig:application1}. 
Note that this idea is similar to transferring 2D-space editing to 3D models~\cite{zhou2005large, orito2020camera}.

\begin{figure}[t]
\centering
\includegraphics[width=\linewidth]{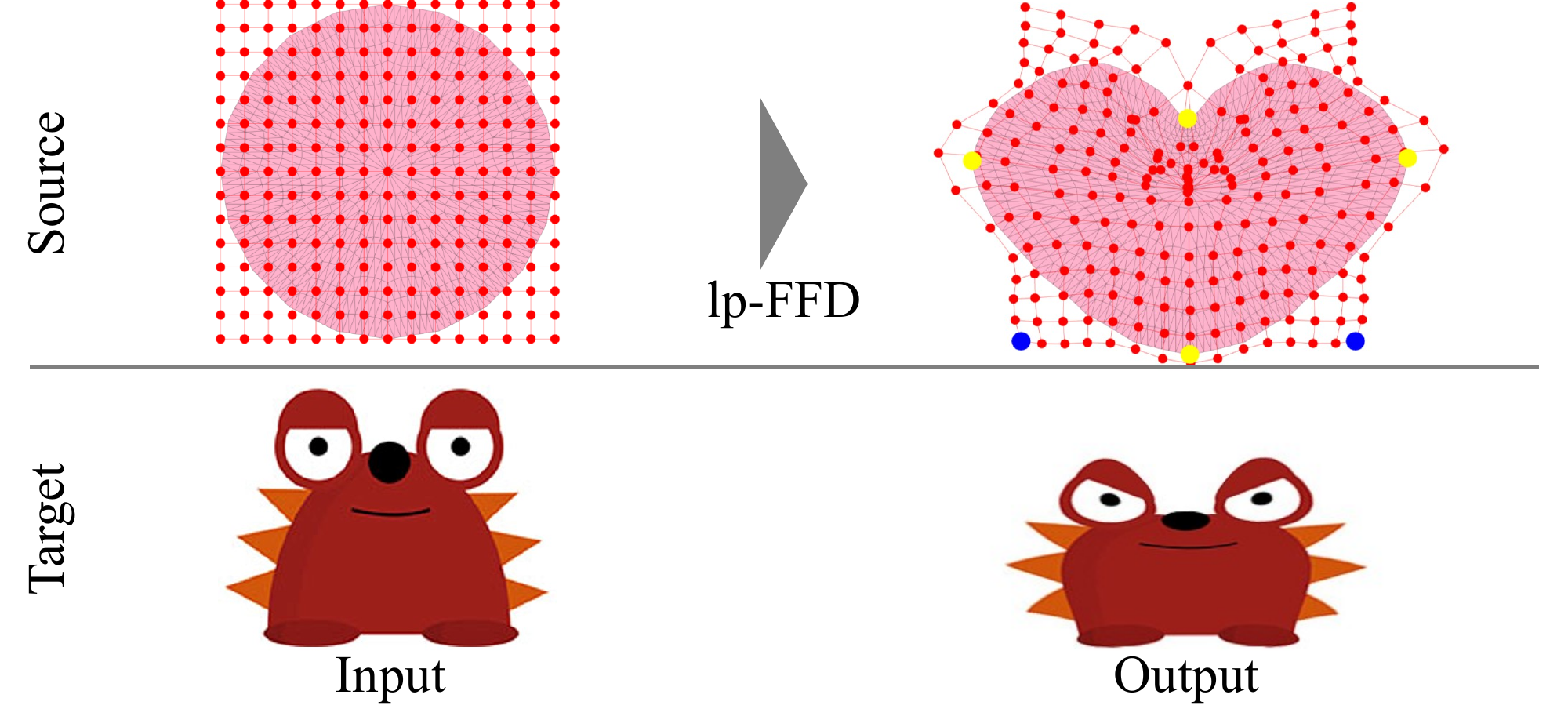}
\caption{An example of deformation transfer. The deformation policy of the source mesh model computed by our lp-FFD (i.e., the deformed grid: $\#P = 15 \times 15$) can easily be reused across different objects (e.g., character image). The character image is from \cite{willett2017triggering}.}
\label{fig:application2}
\end{figure} 

We also think that our lp-FFD can reduce the effort of consumate artists to perform the repetitive deformation work. The DGPs are generally model-specific~\cite{jacobson2011bounded}, meaning that the effort put into obtaining a desired deformation for one model cannot be transferred to make a similar deformation for other objects (e.g., varying geometries or images). In contrast, our lp-FFD enables users to not only perform both direct and indirect manipulations, but also reuse the deformation effects easily since the system's outputs are locations of grid handles (see \autoref{fig:application2}), as with \cite{ju2008reusable, chen2010cage, Li2021auto}. 

In the future, we will take this opportunity to branch out our lp-FFD into the artist's field. Our core idea can be applied to not only 2D/3D FFDs but also to other deformers such as cage-based~\cite{le2017cage} and radial-basis function-based types~\cite{noh2000animated} since they are indirect methods which embed all vertexes of input models into a subspace.
\section{Limitations and Future Work}
\label{sec:limitation}
\begin{figure}[t]
\centering
\includegraphics[width=\linewidth]{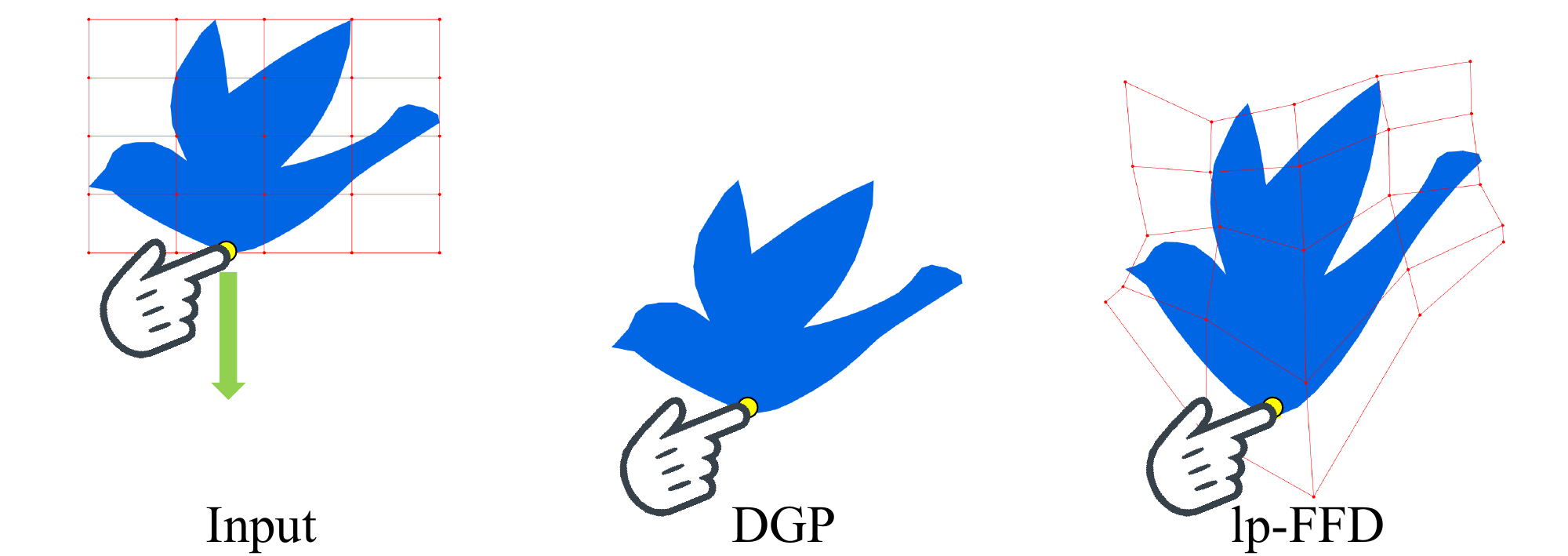}
\caption{Failure case. When manipulating a single-layered model (blue bird: $\#V = 204$) using one vertex handle only, the result with lp-FFD ($\#P = 5\!\times\!5$) is partially stretched, unlike with DGP.}
\label{fig:failure}
\end{figure}

Although our lp-FFD is suitable for deforming input models consisting of multiple isolated meshes, it loses rotational invariance due to $E_{gr}$. Note that this cannot be solved by replacing $E_{gr}$ with the smoothing term in Noh et al.~\cite{noh2021inverse}.
In addition, when only one mesh handle is placed, lp-FFD outputs a partially-stretched shape, while DGP outputs a result as if the input model had been translated rigidly (see \autoref{fig:failure}). 
Therefore, we plan to explore more efficient constraints by rotating and translating the input model based on the user-specified handle locations in the future~\cite{muller2005meshless, muller2016robust}.
%based on the relative location relationships between grid handles like \cite{noh2021inverse}

Since our system does not estimate the rotational information of each grid handle, it remains difficult to apply it to skeletal-based controllers~\cite{jacobson2012fast, orito2020camera}. 
In addition, such controllers require not only handles' information but also model-specific weights computations like \cite{jacobson2011bounded}. 
To address this issue, we plan to extend our idea to automatically optimize the rotation matrix and the weight of each grid handle. 

Existing FFDs are computed by embedding the input model into the pre-defined regular grid. It might be interesting to explore ways to freely resize the grid structure in the deformation step. In addition, we plan to apply our idea to various tasks, such as non-rigid image registration~\cite{sykora2009rigid, Li2021auto} and collision handling for lattice-based dynamics~\cite{rivers2007fastlsm}.

\section{Conclusion}
\label{sec:conc}
This paper has proposed lp-FFD, which allows users to manipulate both vertex handles (i.e., direct manipulation) and grid handles of FFD (i.e., indirect manipulation) while preserving the local details of the input model and making the natural-looking grid shape. 
We have defined a single linear optimizer to solve the locations of grid handles by assigning the FFD formula to a DGP framework. We can easily export the deformed grids in a common format for commercial software, allowing them to be used in existing workflows. There is thus a promising future opportunity to adapt our framework to solve other deformation problems. 

%\section*{Acknowledgments}
%We thank the anonymous reviewers and the editor for their insightful comments, which improved this manuscript. Other acknowledgements removed for review.
%-------------------------------------------------------------------------
% bibtex
\bibliographystyle{eg-alpha-doi} 
\bibliography{main.bib}       

\end{document}